\documentclass[prd,aps,12pt]{revtex4}
\usepackage{graphicx}
\begin{document}

\thispagestyle{empty}
\title{
\hfill{\small {hep-ph/0203199}}\\[2cm]
Baryon magnetic moments and sigma terms in
lattice-regularized chiral perturbation theory}

\author{
Bu\=gra Borasoy$^a$, Randy Lewis$^b$ and Pierre-Philippe A. Ouimet$^{a,b}$}

\affiliation{
$^a$Physik Department, Technische Universit\"at M\"unchen, D-85747 Garching,
    Germany \\
$^b$Department of Physics, University of Regina, Regina, SK, S4S 0A2, Canada
}

\begin{abstract}
\hfil(March 20 2002)
\vspace{15mm}

\hfil{\bf Abstract}

An SU(3) chiral Lagrangian for the lightest decuplet of baryons is
constructed on a discrete lattice of spacetime points, and is added
to an existing lattice Lagrangian for
the lightest octets of mesons and baryons.  A nonzero lattice spacing
renders all loop integrations finite, and the continuum limit of any
physical observable is identical to the result obtained from dimensional
regularization.  Chiral symmetry and gauge invariance
are preserved even at nonzero lattice spacing.  Specific calculations
discussed here include 
the non-renormalization of a conserved vector current, the 
magnetic moments of octet baryons, and the $\pi N$ and $KN$ sigma
terms that relate to the nucleon's strangeness content.
The quantitative difference between physics at a nonzero lattice spacing
and physics in the continuum limit is easily computed, and it represents
an expectation for the size of discretization errors in
corresponding lattice QCD simulations.
\end{abstract}
\maketitle
\newpage

\section{Introduction and discussion}

Chiral perturbation theory (ChPT)\cite{chpt,JM,BKM} is a low momentum effective 
field theory for QCD written as an expansion in small momenta and
quark masses, and it has become an invaluable tool for subatomic physics.
With only the lightest octets of pseudoscalar mesons and spin-1/2 baryons,
ChPT is order-by-order renormalizable and physical results are independent
of whichever regularization prescription is chosen.

The addition of the lightest decuplet of spin-3/2 baryons introduces a
new physical scale, the mass difference between decuplet and octet
baryons, which does not vanish in the chiral limit (i.e. when quark masses
vanish).  However, this mass difference is similar in size to the
pseudoscalar meson masses, so the strict chiral expansion can be
generalized to a ``small scale expansion'' where power counting is now
meaningful even if the decuplet baryons are present.\cite{JM,HHK}

Although physical results are independent of regularization scheme, 
sometimes one scheme is preferable over another such as when
issues other than experimentally-observable quantities are of interest.
For example, the convergence of the ChPT expression for an observable
is often evaluated by comparing the relative sizes of the
contributions that occur at successive orders in the
expansion, even though these individual contributions are not
physically observable.  The work of Refs.~\cite{DHB,BBsigma} discusses this
in some detail, and goes one step further by associating the
renormalization scale with the physical scale of baryon substructure.

For another example of important non-observables, consider lattice QCD.
Numerical simulations must be performed at nonzero lattice spacing, and
computed
results therefore differ from the desired continuum values.  One might
expect that the
effective theory for a discretization of QCD is a discretization of ChPT, i.e.
the most general effective theory, written in terms of hadronic degrees
of freedom, that respects chiral symmetry and the
other symmetries of discretized QCD and that exists in the same discretized
spacetime.  Our previous work, Ref.~\cite{lchpt}, provides one particular
discretization of ChPT along with explicit calculations of differences
between results in this theory and results in the continuum.  

The present work is mainly intended to exemplify the technique of lattice
regularization (with explicit decuplet fields), and to note some of its
features.
Within a chosen lattice ChPT, it is a simple matter to rigorously
determine extrapolations in both quark mass and lattice spacing.
Although less familiar than dimensional regularization,
lattice regularization has the advantage
that loops integrals can easily be performed numerically, since it is a
4-dimensional theory.
The size of lattice spacing artifacts depends on the particular
observable that is studied; at lattice spacings typical of lattice
QCD simulations, we find that our lattice ChPT Lagrangian leads to
noticeable discretization effects for the magnetic moments of octet
baryons but negligible discretization effects for the sigma terms.

It is important to interpret our numerical results appropriately.
If one is interested in using a lattice chiral Lagrangian to rigorously
determine the lattice spacing effects that are present in a specific lattice
QCD simulation, then it will be important to use the chiral Lagrangian that
properly corresponds to whatever lattice QCD action was employed.
Rupak and Shoresh\cite{RandS} provide a list of five $O(a)$ terms that can
appear in the
meson chiral Lagrangian appropriate for a Wilson-type lattice QCD action.
($a$ denotes the lattice spacing.)
Each of these terms is multiplied by its own parameter which depends on the
particular lattice QCD action of interest.  For example, all five of the
coefficients can be made to vanish by using an ``improved'' lattice QCD action.
A similar study could be performed for the baryon Lagrangian.
Our present work does not refer to any specific lattice QCD data,
and this leaves the parameters at $O(a)$ and beyond undetermined.  These
parameters get specified implicitly by our choice of a minimal Lagrangian
where the only lattice spacing effects are contained within simple covariant
derivatives and field representations.  The numerical results for magnetic
moments and sigma terms presented in this work are therefore simply examples
of lattice spacing effects that arise from a Lagrangian which has not been
specially improved in any way.  The main point of our present work is to
demonstrate the use of lattice regularization for chiral Lagrangian
calculations, but we wish to emphasize that this regularization technique can
certainly be applied to extensions of our minimal Lagrangian if one wishes
to fix the lattice spacing parameters to a particular lattice QCD action.

In Sec.~\ref{sec2} of the present work, the Lagrangian of
Ref.~\cite{lchpt} is extended to include the decuplet of spin-3/2 baryons.
Section \ref{sec3} contains a discussion of the electromagnetic vertex:
the non-renormalization of this vertex at vanishing momentum transfer is
shown analytically, and the octet baryon magnetic moments are calculated
as a function of lattice spacing.  In Sec.~\ref{sec4}, the $KN$ sigma terms
of the nucleon's scalar vertex are determined as a function of lattice spacing,
with the necessary chiral
counterterms determined from known baryon masses and the known $\pi N$ sigma
term.  The running of the $\pi N$ sigma term from $q^2=0$ to the Cheng-Dashen
point is also determined as a function of lattice spacing.

\section{A lattice chiral Lagrangian}\label{sec2}

The standard SU(3) chiral Lagrangian containing pseudoscalar mesons (M),
spin-1/2 baryons (B) and spin-3/2 baryons (T) is
\begin{equation}
{\cal L} = {\cal L}_{\rm M} + {\cal L}_{\rm MB} + {\cal L}_{\rm MT}
         + {\cal L}_{\rm MBT}.
\end{equation}
To expand in powers of external momenta, of meson masses and of the T-B mass
difference (relative to the larger scales: baryon masses and the chiral
symmetry breaking scale $\Lambda_\chi\sim 4\pi F_\pi$), it is standard
practice to write the Lagrangian
in terms of heavy baryon fields, $B_v(x)$ and $T_{v\mu}(x)$, instead of the
relativistic fields, $B(x)$ and $T_\mu(x)$.  The transformation for the
spin-1/2 field is
\begin{equation}
B_v(x) = \exp(im_{\rm HB}v\cdot{x})\frac{1}{2}(1+v\!\!\!/)B(x),
\end{equation}
with $m_{\rm HB}$ chosen near the average octet baryon mass.
A similar transformation is used to define the decuplet field\cite{HHK}.
The expanded Lagrangian becomes
\begin{equation}\label{Lexpanded}
{\cal L} = {\cal L}_{\rm M}^{(2)} + {\cal L}_{\rm MB}^{(0)}
         + {\cal L}_{\rm MB}^{(1)} + {\cal L}_{\rm MB}^{(2)}
         + {\cal L}_{\rm MB}^{(3)} + {\cal L}_{\rm MT}^{(1)}
         + {\cal L}_{\rm MBT}^{(1)} + \ldots
\end{equation}
where the omitted terms only contribute to octet baryon properties beyond
third order.
In Euclidean spacetime,
\begin{eqnarray}
{\cal L}_{\rm M}^{(2)}  &=& \frac{F^2}{4}{\rm Tr}\left(\sum_\mu
   \nabla_\mu{U}^\dagger\nabla_\mu{U}-\chi^\dagger{U}-\chi{U}^\dagger\right),
   \label{LM2} \\
{\mathcal{L}}_{\rm MB}^{(0)} &=& (m_0-m_{\rm HB}){\rm Tr}\left(\bar{B}_v
 B_v\right), \\
{\cal L}_{\rm MB}^{(1)} &=& \sum_\mu \bigg[{\rm Tr}\left(\bar{B}_vv_\mu
 D_\mu B_v\right)+{\mathcal{D}}{\rm Tr}\left(\bar{B}_vS_\mu\{u_\mu,B_v\}\right)
+{\cal F}{\rm Tr}\left(\bar{B}_vS_\mu[u_\mu,B_v]\right)\bigg], \\
{\cal L}_{\rm MB}^{(2)} &=& 
\frac{1}{2m_0}{\rm Tr}\left(\bar{B}_v(v\cdot{D}v\cdot{D}-D^2)B_v\right)
-b_{\mathcal{D}}{\rm Tr}\left(\bar{B}_v\{\chi_+,B_v\}\right)
-b_{\mathcal{F}}{\rm Tr}\left(\bar{B}_v[\chi_+,B_v]\right) \nonumber \\
&&-b_0{\rm Tr}\left(\bar{B}_vB_v\right){\rm Tr}\left(\chi_+\right)
 +\frac{i\mu_D}{4m_0}\sum_{\mu,\nu}{\rm Tr}\left(\bar{B}_v[S_\mu,S_\nu]
 \{\xi F_{\mu\nu}^L\xi^\dagger+\xi^\dagger F_{\mu\nu}^R\xi,B_v\}\right)
 \nonumber \\
&& +\frac{i\mu_F}{4m_0}\sum_{\mu,\nu}{\rm Tr}\left(\bar{B}_v[S_\mu,S_\nu]
 [\xi F_{\mu\nu}^L\xi^\dagger+\xi^\dagger F_{\mu\nu}^R\xi,B_v]\right)
+\ldots, \\
{\cal L}_{\rm MB}^{(3)} &=& \ldots, \\
{\cal L}_{\rm MT}^{(1)} &=& -\sum_{\mu\nu}\sum_{ijk} \left(\bar{T}_{v\mu}^{ijk}
v_\nu D_\nu T_{v\mu}^{ijk}-{\cal H}\sum_l\bar{T}_{v\mu}^{ijk}S_\nu u_\nu^{kl}
T_{v\mu}^{ijl}\right)
- \Delta\sum_\mu\sum_{ijk}\bar{T}_{v\mu}^{ijk}T_{v\mu}^{ijk}, \label{LMT} \\
{\cal L}_{\rm MBT}^{(1)} &=& \frac{\cal C}{2}\sum_\mu\sum_{ijklm}
\epsilon_{ijk}\left(\bar{T}_{v\mu}^{klm}u_\mu^{mj}B_v^{li}
+\bar{B}_v^{il}u_\mu^{jm}T_\mu^{klm}\right), \label{LMBT}
\end{eqnarray}
where $S_\mu=\frac{i}{2}\gamma_5\sum_\nu\sigma_{\mu\nu}v_\nu$ is the
Pauli-Lubanski spin vector, $\Delta$ is the decuplet-octet mass difference,
and we choose $v_\mu=(0,0,0,1)$ so that the covariant derivatives of
Ref.~\cite{lchpt} can be retained.
The field strength tensors $F_{\mu\nu}^L(x)$ and $F_{\mu\nu}^R(x)$
correspond to external spin-1 fields.
The superscripted indices in Eqs.~(\ref{LMT}) and (\ref{LMBT}) are flavour
indices:
\begin{eqnarray}
\left(
\matrix{\pi^{11} & \pi^{12} & \pi^{13} \nonumber \\
        \pi^{21} & \pi^{22} & \pi^{23} \nonumber \\
        \pi^{31} & \pi^{32} & \pi^{33} \nonumber   }
\!\!\!\!\!\!\!\!\!\!\!\!\!\!\!\!\! \right)
= \left(
\matrix{{1\over\sqrt 2}\pi^0 + {1\over\sqrt 6}\eta & \pi^+ & K^+ \nonumber \\
        \pi^- & -{1\over\sqrt 2}\pi^0 + {1\over\sqrt 6}\eta & K^0 \nonumber \\
        K^- & \bar{K}^0 & -{2\over\sqrt 6}\eta \nonumber \\}
\!\!\!\!\!\!\!\!\!\!\!\!\!\!\!\!\! \right),
\\
\left(
\matrix{B_v^{11} & B_v^{12} & B_v^{13} \nonumber \\
        B_v^{21} & B_v^{22} & B_v^{23} \nonumber \\
        B_v^{31} & B_v^{32} & B_v^{33} \nonumber   }
\!\!\!\!\!\!\!\!\!\!\!\!\!\!\!\!\! \right)
= \left(
\matrix{{1\over\sqrt 2}\Sigma^0_v + {1\over\sqrt 6}\Lambda_v
        & \Sigma^+_v & p_v \nonumber \\
        \Sigma^-_v & -{1\over\sqrt 2}\Sigma^0_v + {1\over\sqrt 6}\Lambda_v
        & n_v \nonumber \\
        \Xi^-_v & \Xi^0_v & -{2\over\sqrt 6}\Lambda_v \nonumber \\}
\!\!\!\!\!\!\!\!\!\!\!\!\!\!\!\!\! \right),
\\
\left(
\matrix{T_{v\mu}^{111} & T_{v\mu}^{112} & T_{v\mu}^{122} & T_{v\mu}^{222}
        \nonumber \\
      & T_{v\mu}^{113} & T_{v\mu}^{123} & T_{v\mu}^{223} \nonumber \\
      &           & T_{v\mu}^{133} & T_{v\mu}^{233} \nonumber \\
      &           &           & T_{v\mu}^{333} \nonumber \\}
\!\!\!\!\!\!\!\!\!\!\!\!\!\!\!\!\! \right)
= \left(
\matrix{\Delta^{++}_{v\mu} & \frac{1}{\sqrt 3}\Delta^+_{v\mu} &
\frac{1}{\sqrt 3}\Delta^0_{v\mu} & \Delta^-_{v\mu} \nonumber \\
& \frac{1}{\sqrt 3}\Sigma^{*+}_{v\mu} & \frac{1}{\sqrt 6}\Sigma^{*0}_{v\mu} &
\frac{1}{\sqrt 3}\Sigma^{*-}_{v\mu} \nonumber \\
& & \frac{1}{\sqrt 3}\Xi^{*0}_{v\mu} & \frac{1}{\sqrt 3}\Xi^{*-}_{v\mu}
\nonumber \\
& & & \Omega^-_{v\mu} \nonumber \\}
\!\!\!\!\!\!\!\!\!\!\!\!\!\!\!\!\! \right),
\end{eqnarray}
where $T_{v\mu}^{ijk}$ is understood to be completely symmetric in $i,j,k$.
In the Lagrangian, the pseudoscalar mesons are represented nonlinearly,
\begin{equation}
U(x) = \xi^2(x) = \exp(-i\lambda^a\pi^a(x)/F),
\end{equation}
where $\lambda^a$ is a Gell-Mann matrix, and
the current quark mass matrix, $\cal M$, enters via 
\begin{eqnarray}
\chi &=& 2B{\cal M}, \\
\chi_+ &=& \xi^\dagger\chi\xi^\dagger + \xi\chi^\dagger\xi.
\end{eqnarray}

So far the definitions of this section have been general, and have not
assumed a spacetime lattice at all.  We now consider the derivative structures
in the Lagrangian, and it is here that the expressions become lattice
dependent.  As in Ref.~\cite{lchpt}, we choose an isotropic hypercubic
lattice.  Nearest neighbour sites are separated by a distance $a$, and $a_\mu$
will be used to denote a vector of length $a$ in the positive $\mu$ direction.

First, we recall the non-decuplet definitions from Ref.~\cite{lchpt}.
Denoting the external spin-1 fields corresponding to $F_{\mu\nu}^L(x)$
and $F_{\mu\nu}^R(x)$ by $L_\mu(x)$ and $R_\mu(x)$ respectively,
the derivatives that appear in the lowest-order meson Lagrangian,
Eq.~(\ref{LM2}), are taken to be
\begin{equation}
\nabla_\mu^{(+)}U(x) = \frac{1}{a}\left[R_\mu(x)U(x+a_\mu)L_\mu^\dagger(x)
                     - U(x)\right],
\end{equation}
and the meson derivative needed for the baryon Lagrangian is
\begin{equation}
u_\mu(x) = \frac{i}{2}\xi^\dagger(x)\nabla_\mu^{(\pm)}U(x)\xi^\dagger(x)
         - \frac{i}{2}\xi(x)\nabla_\mu^{(\pm)}U^\dagger(x)\xi(x),
\end{equation}
where
\begin{equation}
\nabla_\mu^{(\pm)}U(x) = \frac{1}{2a}\left[R_\mu(x)U(x+a_\mu)L_\mu^\dagger(x)
                       - R^\dagger_\mu(x)U(x-a_\mu)L_\mu(x-a_\mu)\right].
\end{equation}
The covariant derivative for the octet baryon field is
\begin{eqnarray}
aD_4B_v(x) &=& B_v(x) - \sum_{X,Y=L,R}\Gamma^{X\dagger}_4(x-a_4)
                        B_v(x-a_4)\Gamma^Y_4(x-a_4), \\
aD_jB_v(x) &=& \frac{1}{2}\sum_{X,Y=L,R}
               \left[\Gamma^X_j(x)B_v(x+a_j)\Gamma^{Y\dagger}_4(x)
                    -\Gamma^{X\dagger}_j(x-a_j)B_v(x-a_j)\Gamma^Y_4(x-a_j)
               \right], \nonumber \\
\end{eqnarray}
in the temporal and spatial directions respectively, where
\begin{equation}
2\Gamma^X_\mu(x) = \left\{
\begin{array}{ll}
\xi(x)L_\mu(x)\xi^\dagger(x+a_\mu), & {\rm ~for~} X=L, \\
\xi^\dagger(x)R_\mu(x)\xi(x+a_\mu), & {\rm ~for~} X=R.
\end{array}
\right.
\end{equation}

Not defined explicitly in Ref.~\cite{lchpt} but required for the present work
are the field strength tensors, which we discretize as follows so that the
chiral transformation properties and the lattice symmetries are respected:
\begin{eqnarray}
4ia^2F_{\mu\nu}^X(x)&=&4 - X_\mu(x)X_\nu(x+a_\mu)X_\mu^\dagger(x+a_\nu)
                                  X_\nu^\dagger(x) \nonumber \\
                   &&- X_\nu(x)X_\mu^\dagger(x-a_\mu+a_\nu)
                       X_\nu^\dagger(x-a_\nu)X_\mu(x-a_\mu) \nonumber \\
                   &&- X_\mu(x-a_\mu)X_\nu^\dagger(x-a_\mu-a_\nu)
                       X_\mu^\dagger(x-a_\mu-a_\nu)X_\nu(x-a_\nu) \nonumber \\
                   &&- X_\nu^\dagger(x-a_\nu)X_\mu(x-a_\nu)
                       X_\nu(x+a_\mu-a_\nu)X_\mu^\dagger(x),
\end{eqnarray}
for $X=L$ and $R$.
Finally, consider the covariant derivative of the decuplet field.
Since we have chosen $v_\mu=(0,0,0,1)$, only the temporal derivative is
needed and we define it to be
\begin{eqnarray}
aD_4T_{v\mu}^{ijk}(x) &=& T_{v\mu}^{ijk}(x)
                     - \sum_l\left[\Gamma_{4il}^L(x-a_4)
                       +\Gamma_{4il}^R(x-a_4)\right]T_{v\mu}^{ljk}(x-a_4)
                     \nonumber \\
                  && - \sum_l\left[\Gamma_{4jl}^L(x-a_4)
                       +\Gamma_{4jl}^R(x-a_4)\right]T_{v\mu}^{ilk}(x-a_4)
                     \nonumber \\
                  && - \sum_l\left[\Gamma_{4kl}^L(x-a_4)
                       +\Gamma_{4kl}^R(x-a_4)\right]T_{v\mu}^{ijl}(x-a_4).
\end{eqnarray}
The chiral invariance of this Lagrangian can readily be verified by
adding the chiral transformation of the decuplet field,
$T_{v\mu}^{ijk} \to o_{il}o_{jm}o_{kn}T_{v\mu}^{lmn}$, to the
set of chiral transformations given in Ref.~\cite{lchpt}.

When performing calculations, it is important to remember that the
decuplet propagator must always be accompanied by the projector which
eliminates spurious spin-1/2 components from the propagating field.\cite{HHK}
For the present Lagangian, the product of propagator and projector is
\begin{equation}
\frac{-P^{3/2}_{\mu\nu}}{\Gamma_{TT}}
 = \frac{ia(\delta_{\mu\nu}-v_\mu v_\nu+\frac{4}{3}
         S_\mu S_\nu)}{\sin(ap_4)-i[a\Delta+2\sin^2(ap_4/2)]}.
\end{equation}

\section{Gauge invariance and octet baryon magnetic moments}\label{sec3}

To calculate the electromagnetic form factors of an octet baryon,
one simply identifies the photon field, ${\cal A}_\mu(x)$,
with the external spin-1 fields of Sec.~\ref{sec2},
\begin{equation}
L_\mu(x) = R_\mu(x) = \exp\left[-iaeQ{\cal A}_\mu(x)\right],
\end{equation}
where $Q=diag(2/3,-1/3,-1/3)$.

\begin{figure}[tbh]
\includegraphics[height=12.0cm]{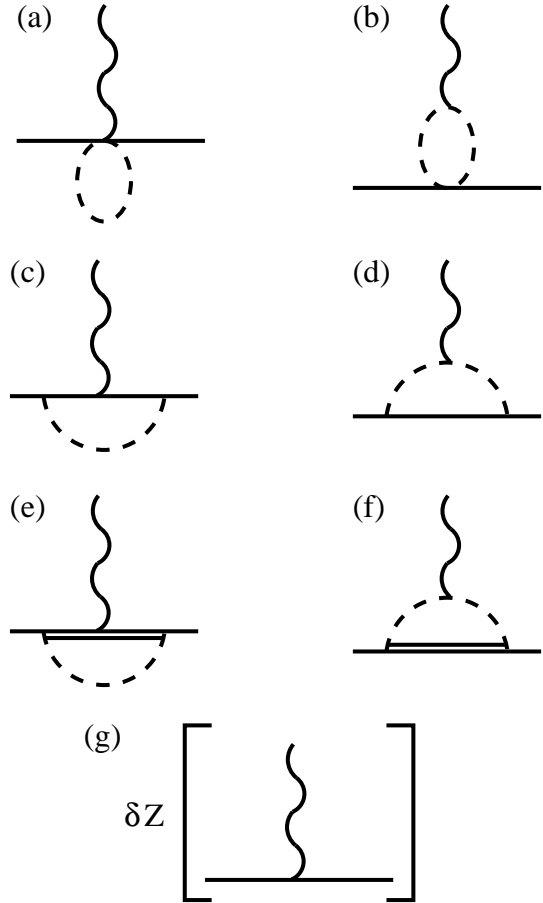}
\caption{One-loop contributions to an octet baryon's electromagnetic
         vertex.  Dashed, solid, double and wavy lines represent
         mesons, octet baryons, decuplet baryons and photons
         respectively.  $\delta Z$ denotes the contribution to
         wave function renormalization that arises from the diagrams
         in Fig.~\protect\ref{fig:wavefn}.
        }\label{fig:EMvertex}
\end{figure}

{}From Eqs.~(\ref{Lexpanded}-\ref{LMBT}), it is easy to see that the
leading order contribution to an octet baryon's electromagnetic vertex
is contained within ${\cal L}_{\rm MB}^{(1)}$ and the next-to-leading
order contribution within ${\cal L}_{\rm MB}^{(2)}$.
The resulting matrix element is
\begin{eqnarray}\label{EMampl}
\left< B(p^\prime)\left|J_\mu^{\rm em}\right|B(p)\right>
&=& ie\bar{B}_v(k^\prime)\bigg[
v_\mu Q_B\left(1-i\sin(ak^\prime_4)-2\sin^2\left(\frac{ak^\prime_4}
{2}\right)\right) \nonumber \\
&+& 
 \frac{i\mu_B^{\rm LO}}{am_0}\sum_\nu[S_\mu,S_\nu]\sin(aq_\nu)
 \left(1-\frac{i}{2}\sin(aq_\mu)-\sin^2\left(\frac{aq_\mu}{2}\right)\right)
\bigg]B_v(k),
\end{eqnarray}
where  $q=p^\prime-p$ is the momentum transfer,
$Q_B$ denotes the electric charge of the baryon, and
$\mu_B^{\rm LO}$ are the leading order expressions for the magnetic
moments:
\begin{eqnarray}
\mu_p^{\rm LO} = \mu_{\Sigma^+}^{\rm LO} &=& \frac{\mu_D}{3} + \mu_F, \\
\mu_{\Sigma^-}^{\rm LO} = \mu_{\Xi^-}^{\rm LO} &=& \frac{\mu_D}{3} - \mu_F, \\
\mu_{\Sigma^0}^{\rm LO} = -\mu_{\Lambda}^{\rm LO} &=& \frac{\mu_D}{3}, \\
\mu_n^{\rm LO} = \mu_{\Xi^0}^{\rm LO} &=& -\frac{2\mu_D}{3}, \\
\mu_{\Lambda\Sigma^0}^{\rm LO} &=& \frac{\mu_D}{\sqrt{3}}.
\end{eqnarray}
Notice from Eq.~(\ref{EMampl}) that the term containing $Q_B$ is purely
temporal and the $\mu_B^{\rm LO}$ term is purely spatial with our
chosen frame, since $v_\mu=(0,0,0,1) \Rightarrow S\cdot{v}=0$.

Returning to the full Lagrangian of Eq.~(\ref{Lexpanded}),
one finds the corrections to the matrix element of Eq.~(\ref{EMampl}) 
that are shown diagrammatically in Fig.~\ref{fig:EMvertex}.
As will now be shown, evaluation of these diagrams leads to the
renormalization of $\mu_B^{\rm LO}$, and to the non-renormalization of
$Q_B$ as required by gauge invariance.

\subsection{The non-renormalization of electric charge}

To verify the non-renormalization of $Q_B$, it is sufficient to work
at vanishing momentum transfer.  Also, the momentum of each external
baryon is simply $(0,0,0,m_B)$ plus corrections which are of negligibly
high order in the chiral expansion.  For definiteness we will
discuss the proton's charge in this subsection; the extension
to other octet baryons is straightforward.

\begin{figure}[tbh]
\includegraphics[height=2.0cm]{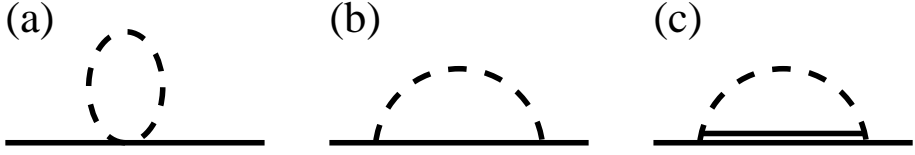}
\caption{One-loop contributions to the wave function renormalization
         of an octet baryon.
         Dashed, solid and double lines represent
         mesons, octet baryons and decuplet baryons respectively.
        }\label{fig:wavefn}
\end{figure}
In this limit,
the contribution of the diagram in Fig.~\ref{fig:EMvertex}(a) to
the matrix element in Eq.~(\ref{EMampl}) with a $\pi^+$ in the loop is
\begin{eqnarray}
{\cal M}_{\pi^+}^{(a)} &=& \int_{-\pi/a}^{\pi/a}\frac{{\rm d}^4q}{(2\pi)^4}
\left(\frac{-iev_\mu
}{2F^2}\right)\left(\frac{a^2}{2\sum_\lambda[1-\cos(aq_\lambda)]
+a^2x_\pi^2}\right) \nonumber \\
&=& \frac{-ia^2ev_\mu}{4F^2}\int_0^\infty{\rm d}z\int_{-\pi/a}^{\pi/a}
    \frac{{\rm d}^4q}{(2\pi)^4}\exp\left[-z\left(4+\frac{a^2x_\pi^2}{2}
    -\sum_\lambda\cos(aq_\lambda)\right)\right] \nonumber \\
&=& \frac{-iev_\mu}{4a^2F^2}W_4(a^2x_\pi^2),
\end{eqnarray}
where $x_M$ is related to the meson mass\cite{lchpt}
\begin{equation}
m_M = \frac{2}{a}{\rm arcsinh}\left(\frac{ax_M}{2}\right),
\end{equation}
and $W_4(\epsilon^2)$ is the integral transform of the fourth power of a Bessel
function\cite{lchpt},
\begin{equation}
W_n(\epsilon^2) \equiv \int_0^\infty{\rm d}x\,I_0^n(x)\exp\left[-x\left(n+
\frac{\epsilon^2}{2}\right)\right].
\end{equation}
Summing over all internal mesons gives
\begin{eqnarray}\label{Ma}
{\cal M}^{(a)} \equiv \sum_M{\cal M}_M^{(a)} &=&
\frac{-iev_\mu}{4a^2F^2}\left[\frac{7}{12}+\left(1-\frac{a^2x_\pi^2}{16}\right)
                       W_4(a^2x_\pi^2)
                       +2\left(1-\frac{a^2x_K^2}{16}\right)W_4(a^2x_K^2)
                       \right.\nonumber \\
     && \left.-\frac{5a^2x_\eta^2}{48}W_4(a^2x_\eta^2) \right].
\end{eqnarray}
Notice that this contribution is quadratically divergent as $a\to0$.
Dimensional regularization does not show power divergences, and
Fig.~\ref{fig:EMvertex}(a) vanishes exactly in that scheme.

Figure \ref{fig:EMvertex}(b) is nonzero only if the loop meson is 
$\pi^\pm$ or $K^\pm$.
For the pion loop, the lattice regularized expression is
\begin{eqnarray}
{\cal M}_{\pi^+}^{(b)} &=& \frac{iea^2v_\mu}{4F^2}
\int_{-\pi/a}^{\pi/a}\frac{{\rm d}^4q}{(2\pi)^4}
\frac{\sin^2(aq_4)}{\left[4+a^2x_\pi^2/2-\sum_\lambda\cos(aq_\lambda)
\right]^2} \nonumber \\
&=& \frac{iev_\mu}{4a^2F^2}\int_0^\infty{\rm d}z\,I_0^3(z)I_0^\prime(z)
    \exp\left[-z\left(4+\frac{a^2x_\pi^2}{2}\right)\right] \nonumber \\
&=& \frac{-iev_\mu}{4a^2F^2}\left[\frac{1}{4}-\left(1+\frac{a^2x_\pi^2}{8}
    \right)W_4(a^2x_\pi^2)\right].
\end{eqnarray}
Repeating this calculation for the kaon loop leads to
\begin{equation}\label{Mb}
{\cal M}^{(b)} \equiv \sum_M{\cal M}_M^{(b)} =
\frac{-iev_\mu}{4a^2F^2}\left[\frac{3}{4}\
                -\left(1+\frac{a^2x_\pi^2}{8}\right)W_4(a^2x_\pi^2)
                -2\left(1+\frac{a^2x_K^2}{8}\right)W_4(a^2x_K^2)\right].
\end{equation}
The only other contribution to the renormalization of $Q_B$ that is
independent of the axial couplings ($\cal D$ and $\cal F$) and the
decuplet (with coupling $\cal C$) comes from
the tadpole contribution to wavefunction renormalization.
As discussed in Ref.~\cite{lchpt}, the lowest-order octet baryon
two-point function is
\begin{equation}
\Gamma_{BB} = m_{\rm HB} - m_0
                  -\frac{i}{a}\sum_\mu{v_\mu}\left[\sin(aq_\mu)
                  -2i\sin^2\left(\frac{aq_\mu}{2}\right)\right].
\end{equation}
The one-loop correction shown in Fig.~\ref{fig:wavefn}(a) evaluates as
follows:
\begin{equation}
\delta\Gamma_{pp}^{(a)} =
\left[\frac{-i}{a}\left(\sin(ak_4)-2i\sin^2\left(\frac{ak_4}{2}\right)\right)
+\frac{1}{a}\right]\frac{1}{\delta Z_p^{(a)}},
\end{equation}
where the external proton momentum is $m_0v+k$ and
\begin{equation}\label{Zp}
\delta Z_p^{(a)} =
             \frac{1}{3a^2F^2}\left[1-\frac{9}{64}a^2x_\pi^2W_4(a^2x_\pi^2)
            -\frac{9}{32}a^2x_K^2W_4(a^2x_K^2)
            -\frac{5}{64}a^2x_\eta^2W_4(a^2x_\eta^2)\right].
\end{equation}
This is precisely what was required to facilitate the expected
non-renormalization, since Eqs.~(\ref{Ma}), (\ref{Mb}) and (\ref{Zp}) give
\begin{equation}
{\cal M}^{(a)} + {\cal M}^{(b)} + iev_\mu\delta Z_p^{(a)} = 0.
\end{equation}

Turning to the contributions that depend on $\cal D$ and $\cal F$,
one finds
\begin{eqnarray}\label{Mcpi0}
{\cal M}^{(c)}_{\pi^0} &=& \frac{1}{2}({\cal D}+{\cal F})^2G_0(a^2x_\pi^2), \\
{\cal M}^{(c)}_{K^0} &=& ({\cal D}-{\cal F})^2G_0(a^2x_K^2), \label{McK0} \\
{\cal M}^{(c)}_\eta &=& \frac{1}{6}({\cal D}-3{\cal F})^2G_0(a^2x_\eta^2),
\label{Mceta}
\end{eqnarray}
where
\begin{equation}
G_M(\epsilon^2) = \frac{iea^2v_\mu}{4F^2}\int_{-\pi/a}^{\pi/a}\frac{{\rm d}^4q}
              {(2\pi)^4}\frac{[\cos(aq_4)-i\sin(aq_4)]\sum_{k=1}^3\sin^2(aq_k)}
              {[4+\epsilon^2/2-\sum_\lambda\cos(aq_\lambda)]
              [\sin(aq_4)-iaM-2i\sin^2(aq_4/2)]^2}.
\end{equation}
The contribution of Fig.~\ref{fig:EMvertex}(d) with a charged pion in the
loop is
\begin{eqnarray}
{\cal M}^{(d)}_{\pi^+} &=& \frac{iea^2}{F^2}({\cal D}+{\cal F})^2
\int_{-\pi/a}^{\pi/a}\frac{{\rm d}^4q}{(2\pi)^4}
\frac{v_\mu\sin(aq_4)\sum_\rho S_\rho\sin
(aq_\rho)\sum_\sigma S_\sigma\sin(aq_\sigma)}
{[\sin(aq_4)-2i\sin^2(aq_4/2)][4+a^2x_\pi^2/2-\sum_\lambda\cos(aq_\lambda)]}
\nonumber \\
&=& \frac{iea^2v_\mu}{2F^2}({\cal D}+{\cal F})^2\times \nonumber \\
&& \frac{{\rm d}~~}
    {{\rm d}(a^2x_\pi^2)}\int_{-\pi/a}^{\pi/a}\frac{{\rm d}^4q}{(2\pi)^4}
    \frac{\sin(aq_4)\sum_{k=1}^3\sin^2(aq_k)}
{[\sin(aq_4)-2i\sin^2(aq_4/2)][4+a^2x_\pi^2/2-\sum_\lambda\cos(aq_\lambda)]}
\nonumber \\
&=& ({\cal D}+{\cal F})^2G_0(a^2x_\pi^2), \label{Mdpi}
\end{eqnarray}
where we have used $S_4=0$ and $S_1^2=S_2^2=S_3^2=-1/4$ as well as
integration by parts.
Similarly,
\begin{equation}\label{MdK}
{\cal M}^{(d)}_{K^+} = \frac{2}{3}({\cal D}^2+3{\cal F}^2)G_0(a^2x_K^2).
\end{equation}

Next, consider the wave function renormalization of Fig.~\ref{fig:wavefn}(b).
For a $\pi^0$ loop, the contribution to the two-point function is
\begin{eqnarray}
\delta\Gamma_{pp}^{(b)} &=& \frac{-ia}{2F^2}({\cal D}+{\cal F})^2
\int_{-{\rm min}(\pi/a,\pi/a+k)}^{{\rm min}(\pi/a,\pi/a-k)}
\frac{{\rm d}^4q}{(2\pi)^4} \times \nonumber \\
&& \frac{\sum_\rho S_\rho\sin (aq_\rho)\sum_\sigma S_\sigma\sin(aq_\sigma)}
{[\sin(aq_4+ak_4)-2i\sin^2((aq_4+ak_4)/2)]
[4+a^2x_\pi^2/2-\sum_\lambda\cos(aq_\lambda)]}
\nonumber \\
&\equiv& \frac{1}{\delta Z_{p\pi^0}^{(b)}}\left[\frac{\sin(ak_4)-2i\sin^2(ak_4)
         -a\delta X}{ia}\right]
\end{eqnarray}
where the external baryon momentum is $m_0v+k$ and
$\delta X$ produces a mass renormalization.
The limits of integration have been chosen to ensure that the momentum
of each internal propagator remains within the lattice's Brillouin zone
over the entire domain of $q$ integration.\cite{lchpt}
As it happens, the $k$-dependences in the limits of
integration do not affect wave function renormalization, and the result is
\begin{equation}
iev_\mu\delta Z_{p\pi^0}^{(b)} =
       -\frac{1}{2}({\cal D}+{\cal F})^2G_0(a^2x_\pi^2),
\end{equation}
which exactly cancels the $\pi^0$ loop from Eq.~(\ref{Mcpi0}).
In the same way, replacing the $\pi^0$ in $\delta Z$ by each of the
other mesons serves to exactly cancel Eqs.~(\ref{McK0}), (\ref{Mceta}),
(\ref{Mdpi}) and (\ref{MdK}).

Finally, the decuplet contributions of Figs.~\ref{fig:EMvertex}(e),
\ref{fig:EMvertex}(f) and \ref{fig:wavefn}(c) are found to cancel in
essentially that same manner as did the octet contributions:
\begin{eqnarray}
{\cal M}^{(e)}_{\pi^+} &=& \frac{4}{9}{\cal C}^2G_\Delta(a^2x_\pi^2), \\
ie\delta Z^{(c)}_{p\pi^+} &=& -\frac{8}{9}{\cal C}^2G_\Delta(a^2x_\pi^2), \\
{\cal M}^{(e)}_{\pi^0} = -ie\delta Z^{(c)}_{p\pi^0}
                      &=& \frac{4}{9}{\cal C}^2G_\Delta(a^2x_\pi^2), \\
{\cal M}^{(e)}_{K^0} = -ie\delta Z^{(c)}_{K^0} &=&
                          \frac{2}{9}{\cal C}^2G_\Delta(a^2x_K^2), \\
{\cal M}^{(f)}_{\pi^+} &=& -\frac{4}{9}{\cal C}^2G_\Delta(a^2x_\pi^2), \\
{\cal M}^{(f)}_{K^+} = -ie\delta Z^{(c)}_{pK^+}
                         &=& \frac{1}{9}{\cal C}^2G_\Delta(a^2x_K^2).
\end{eqnarray}
Thus, all corrections to the proton's electric charge at first loop order
in the chiral expansion sum to zero, and the non-renormalization of electric
charge is confirmed.

\subsection{The octet baryon magnetic moments}

The one-loop corrections to octet baryon magnetic moments come from
Figs.~\ref{fig:EMvertex}(d) and \ref{fig:EMvertex}(f).  These diagrams
were discussed in the previous subsection, but only for vanishing momentum
transfer which is not sufficient to obtain magnetic moments.

It is convenient to choose the Breit frame where the incoming baryon has
momentum $m_0v-q/2$ and the outgoing baryon has momentum $m_0v+q/2$, and
to choose the integration momentum to be the internal baryon's momentum.
This choice displays symmetries in the integrand that help to simplify
the calculations.  Notice that higher order corrections to the external
baryon momenta have already been omitted, and that $m_0$ is equal to
the physical mass in these loop diagrams since the difference is of higher
chiral order.

The evaluation of Figs.~\ref{fig:EMvertex}(d) and \ref{fig:EMvertex}(f)
for arbitrary momentum transfer $q$ gives
\begin{equation}
{\cal M}^{(d)} + {\cal M}^{(f)} = \frac{-2e}{am_N}\sum_\nu[S_\mu,S_\nu]
\sin\left(\frac{aq_\nu}{2}\right)\delta F_2(q),
\end{equation}
where $\delta F_2$ is the loop correction to the Pauli form factor.
The magnetic moments are
\begin{equation}
\mu_B = \mu_B^{\rm LO}
 + \left[\delta F_2^{\rm oct}(0)\right]_B
 + \left[\delta F_2^{\rm dec}(0)\right]_B,
\end{equation}
where the last two terms are from Figs.~\ref{fig:EMvertex}(d) and
\ref{fig:EMvertex}(f) respectively.
Both of these terms rely on a single integral with $\mu\neq4$, $\nu\neq4$ and
$\mu\neq\nu$, namely
\begin{equation}
H_M(\epsilon^2) = \frac{-ia^3m_N}{2F^2}\int_{-\pi/a}^{\pi/a}\frac{{\rm d}^4q}
 {(2\pi)^4}\frac{\sin^2(aq_\mu)\cos(aq_\nu)}{\left[4+\epsilon^2/2-\sum_\lambda
 \cos(aq_\lambda)\right]^2\left[\sin(aq_4)-iaM-2i\sin^2(aq_4/2)\right]},
\end{equation}
but with one key difference.  For the decuplet diagram it is $H_\Delta$ that
enters, and the mass splitting
$\Delta$ ensures that the integrand has no singularities on the domain
of integration.  In that case, a convenient form for numerical evaluation is
\begin{equation}
H_\Delta(\epsilon^2)
 = \frac{a^3m_N}{4F^2}\int_{-\pi/a}^{\pi/a}\frac{{\rm d}^4q}
 {(2\pi)^4}\frac{\cos(aq_\mu)\cos(aq_\nu)[1+a\Delta-\cos(aq_4)]}
 {\left[4+\epsilon^2/2-\sum_\lambda
 \cos(aq_\lambda)\right]\left[(1+a\Delta)(1-\cos(aq_4))+a^2\Delta^2/2\right]}.
\end{equation}
\begin{table}[thb]
\caption{Coefficients that appear in the octet baryon magnetic
         moments.}\label{tab:coeffs}
\begin{tabular}{ccccccc}
\hline
$B$ & ~~~~~$c_1^B$~~~~~ & ~~~~~$c_2^B$~~~~~ & $c_3^B$ & $c_4^B$ & $c_5^B$
    & $c_6^B$ \\
\hline
$p$ & 1/3 & 1 & $({\cal D}+{\cal F})^2$ & $(2/3){\cal D}^2+2{\cal F}^2$
    & 2/9 & -1/18 \\
$n$ &-2/3 & 0 & $-({\cal D}+{\cal F})^2$ & $({\cal D}-{\cal F})^2$
    &-2/9 & -1/9 \\
$\Sigma^+$ & 1/3 & 1 & $(2/3){\cal D}^2+2{\cal F}^2$ & $({\cal D}+{\cal F})^2$
    &-1/18 & 2/9 \\
$\Sigma^0$ & 1/3 & 0 & 0 & $2{\cal DF}$ & 0 & 1/6 \\
$\Sigma^-$ & 1/3 &-1 & $-(2/3){\cal D}^2-2{\cal F}^2$ & $-({\cal D}-{\cal F})^2$
    & 1/18 & 1/9 \\
$\Xi^0$ &-2/3 & 0 & $({\cal D}-{\cal F})^2$ & $-({\cal D}+{\cal F})^2$
    &-1/9 & -2/9 \\
$\Xi^-$ & 1/3 &-1 & $-({\cal D}-{\cal F})^2$ & $-(2/3){\cal D}^2-2{\cal F}^2$
    & 1/9 & 1/18 \\
$\Lambda$ &-1/3 & 0 & 0 & $-2{\cal DF}$ & 0 & -1/6 \\
$\Lambda\Sigma^0$ & $1/\sqrt{3}$ & 0 & $4/(\sqrt{3}){\cal DF}$
       & $(2/\sqrt{3}){\cal DF}$ & ~~$1/(3\sqrt{3})$~~ & ~~$1/(6\sqrt{3})$~~
\\ \hline
\end{tabular}
\end{table}
For the octet diagram there is a singularity, and
we must integrate around it
according to the usual ``$+i\epsilon$'' prescription for field theory.
Some details of this procedure in the context of lattice regularization
can be found in the appendices of Ref.~\cite{lchpt}, so here we simply
provide our final expression:
\begin{eqnarray}
H_0(\epsilon^2)
&=&\frac{a^3m_N}{4F^2}\left[ \int_{-\pi/a}^{\pi/a}\frac{{\rm d}^4q}
 {(2\pi)^4}\frac{\cos(aq_\mu)\cos(aq_\nu)}
 {\left[4+\epsilon^2/2-\sum_\lambda\cos(aq_\lambda)\right]} \right.\nonumber \\
&&+ \left. \int_{-\pi/a}^{\pi/a}\frac{{\rm d}^3q}
 {(2\pi)^3}\frac{\cos(aq_\mu)\cos(aq_\nu)}
 {\left[3+\epsilon^2/2-\sum_{k=1}^3\cos(aq_k)\right]} \right].
\end{eqnarray}
This too is easily evaluated numerically.

\begin{figure}[tbh]
\vspace*{9mm}
\includegraphics[height=10.0cm]{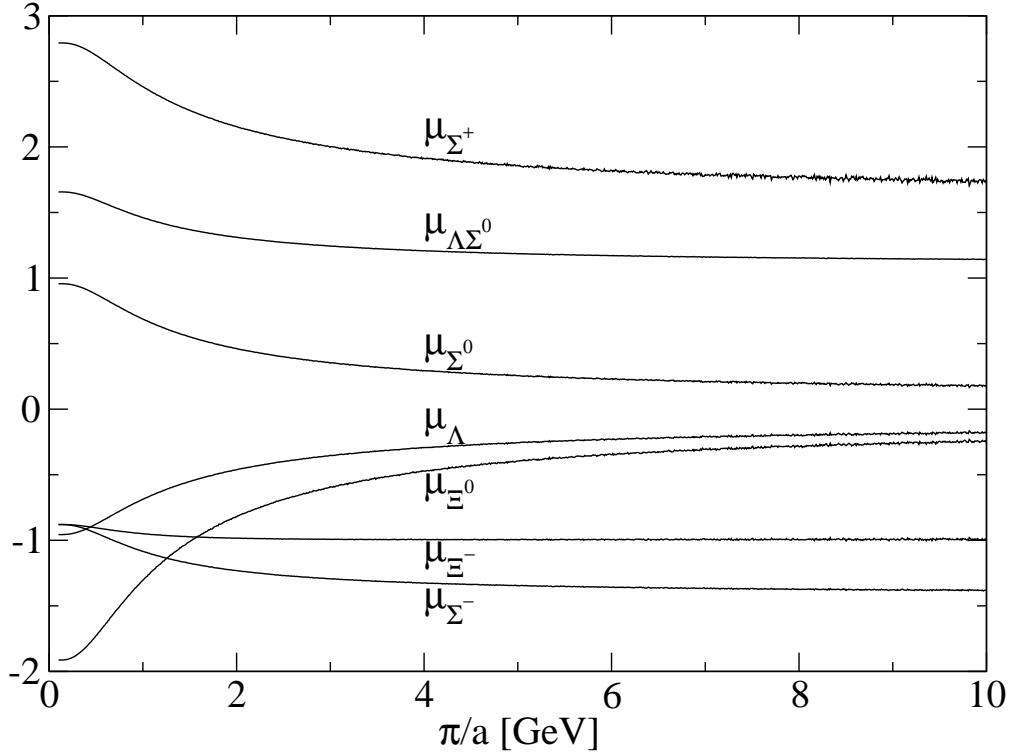}
\caption{Octet baryon magnetic moments (in units of $\mu_N$) as functions
         of lattice spacing.
         Lagrangian parameters are fixed by requiring the proton and
         neutron magnetic moments to equal their experimental values
         at all lattice spacings.
        }\label{fig:muplot}
\end{figure}
Our final expression for the octet baryon magnetic moments is
\begin{equation}
\mu_B = c_1^B\mu_D + c_2^B\mu_F + c_3^BH_0(a^2x_\pi^2) + c_4^BH_0(a^2x_K^2)
                   + c_5^BH_\Delta(a^2x_\pi^2) + c_6^BH_\Delta(a^2x_K^2),
\end{equation}
where the coefficients are listed in Table~\ref{tab:coeffs}.
$H_M$ diverges as $a\to0$, but the offending terms can be absorbed into
renormalized definitions of $\mu_D$ and $\mu_F$.  The result is identical to
that obtained from dimensional regularization.  However lattice
regularization also allows us to compute at nonzero $a$, and it is
interesting to consider lattice spacings that are typical of lattice QCD
simulations.

Using the experimental values of $\mu_p$ and $\mu_n$ to fix the parameters
$\mu_D$ and $\mu_F$, the other seven magnetic moments become predictions
of the theory, and are plotted as a function of the lattice cutoff $\pi/a$
in Fig.~\ref{fig:muplot}.
The plot assumes standard experimental values for the coefficients appearing
within loops: ${\cal D}=0.75$, ${\cal F}=0.50$, and ${\cal C}=1.5$.
These values could be varied within experimental uncertainties, but
such details do not significantly affect our present interest:
the size of discretization effects.
\begin{table}[thb]
\caption{A comparison of the magnetic moments at $\pi/a=6.0$ GeV,
         corresponding to $a=0.10$ fm, and their values in the continuum limit.
        }\label{tab:muratio}
\begin{tabular}{cccccccc}
\hline
$B$ & ~~~~$\Sigma^+$~~~~ & ~~~~$\Sigma^0$~~~~ & ~~~~$\Sigma^-$~~~~ & 
      ~~~~$\Xi^0$~~~~ & ~~~~$\Xi^-$~~~~ & ~~~~~$\Lambda$~~~~~ &
      ~~~$\Lambda\Sigma^0$~~~ \\
\hline
$\mu_B(a=0)$ & 1.64 & 0.12 & -1.40 & -0.14 & -0.98 & -0.12 & 1.11 \\
$\mu_B(a=0.1{\rm fm})/\mu_B(a=0)$ & 1.11 & 1.94 & 0.97 & 2.58 & 1.01
             & 1.95 & 1.05
\\ \hline
\end{tabular}
\end{table}

As seen in Fig.~\ref{fig:muplot}, each magnetic moment
smoothly approaches the corresponding dimensional regularized
result at $\pi/a\to\infty$.  These limiting values are given in
Table~\ref{tab:muratio}.  The agreement with experiment is not particularly 
impressive at this chiral order, as has been known for some time\cite{JLMS}.
The situation is dramatically improved at next chiral order\cite{PRM,KM} or
by the methods of Refs.~\cite{DHB,BBsigma}.
At present, we focus on the discretization effects rather than a precise
comparison to experiment.

For $\pi/a=6$ GeV, Table~\ref{tab:muratio} shows that the
relative sizes of discretization effects vary from a few percent to a factor
of 2 or more, depending on which magnetic moment is chosen.
The large variation is somewhat misleading: the absolute discretizations
are quite comparable for all magnetic moments, as is evident from
Fig.~\ref{fig:muplot}.  However, even $O(10\%\to30\%)$ discretization
uncertainties
are significant in this context, since chiral corrections are typically of
this order too.  If ChPT is being employed as a way to determine chiral
effects in lattice QCD, then these discretization effects must be
considered.
Notice that $\pi/a=6$ GeV corresponds to a lattice spacing of 0.1 fm, which is
typical of modern lattice QCD simulations.

\section{The $\pi N$ and $KN$ sigma terms}\label{sec4}

Sigma terms are the scalar form factors of a baryon multiplied by the
quark mass,
\begin{eqnarray}
\sigma_{\pi N}(t) &=& \hat{m}\left< N(p^\prime)\left|\bar{u}u+\bar{d}d\right|
                      N(p)\right>, \\
\sigma_{KN}^{(1)}(t) &=& \frac{1}{2}(\hat{m}+m_s)\left< N(p^\prime)\left|
                         \bar{u}u+\bar{s}s\right|N(p)\right>, \\
\sigma_{KN}^{(2)}(t) &=& \frac{1}{2}(\hat{m}+m_s)\left< N(p^\prime)\left|
                         -\bar{u}u+2\bar{d}d+\bar{s}s\right|N(p)\right>,
\end{eqnarray}
where $\hat m = (m_u+m_d)/2$.
The sigma terms vanish in the chiral limit and are therefore useful in
discussions
of chiral symmetry and its breaking.  Furthermore, they offer a probe
of the nucleon's strangeness content,
\begin{equation}
y \equiv \frac{2\left< N(p)\left|\bar{s}s\right|N(p)\right>}
           {\left< N(p)\left|\bar{u}u+\bar{d}d\right|N(p)\right>}
  = \left(\frac{\hat m}{\hat m + m_s}\right)
    \left(\frac{3\sigma_{KN}^{(1)}(0)+\sigma_{KN}^{(2)}(0)}{\sigma_{\pi N}(0)}
    \right) - 1,
\end{equation}
and this quantity continues to be of great interest to many
researchers.  (See Ref.~\cite{BBsigma} and references therein.)

The lattice-regularized Lagrangian leads to
\begin{eqnarray}
\sigma_{\pi N} &=& -x_\pi^2(2b_0+b_D+b_F) + x_\pi^2\sum_{j=1}^3
                   \sum_{\phi}a_j^\phi K_j(\phi,\phi), \\
\sigma_{KN}^{(1)} &=& -2x_K^2(b_0+b_D) + x_K^2\sum_{j=1}^3
                   \sum_{\phi}b_j^\phi K_j(\phi,\phi)
       - \frac{x_K^2}{12}({\cal D}^2-2{\cal DF}-3{\cal F}^2)K_1(\pi,\eta)
       \nonumber \\ &&
       + \frac{x_K^2}{6}K_2(\pi,\eta), \\
\sigma_{KN}^{(2)} &=& -2x_K^2(b_0-b_F) + x_K^2\sum_{j=1}^3
                   \sum_{\phi}c_j^\phi K_j(\phi,\phi)
       + \frac{x_K^2}{4}({\cal D}^2-2{\cal DF}-3{\cal F}^2)K_1(\pi,\eta)
       \nonumber \\ &&
       - \frac{x_K^2}{2}K_2(\pi,\eta),
\end{eqnarray}
\begin{table}[thb]
\caption{Coefficients that appear in the nucleon sigma terms.
        }\label{tab:sigcoeffs}
\begin{tabular}{ccccc}
\hline
$j$ & $\phi$ & $a_j^\phi$ & $b_j^\phi$ & $c_j^\phi$ \\
\hline
1 & $\pi$ & $(3/2)({\cal D}+{\cal F})^2$ & $(3/4)({\cal D}+{\cal F})^2$ & 
                                       $(3/4)({\cal D}+{\cal F})^2$ \\
1 & $K$ & $(1/6)(5{\cal D}^2-6{\cal DF}+9{\cal F}^2)$ &
          $(7/6){\cal D}^2-{\cal DF}+(5/2){\cal F}^2$ & 
                                       $(3/2)({\cal D}-{\cal F})^2$ \\
1 & $\eta$ & $(1/18)({\cal D}-3{\cal F})^2$ & $(5/36)({\cal D}-3{\cal F})^2$ & 
                                       $(5/36)({\cal D}-3{\cal F})^2$ \\
2 & $\pi$ & 3/2 & 3/4 & 3/4 \\
2 & $K$ & 3/2 & 5/2 & 3/2 \\
2 & $\eta$ & 3/4 & 3/2 & 25/36 \\
3 & $\pi$ & 2 & 1 & 1 \\
3 & $K$ & 1/4 & 1/3 & 1/2 \\
3 & $\eta$ & 0 & 0 & 0 
\\ \hline
\end{tabular}
\end{table}
where the integrals are
\begin{eqnarray}
K_1(\phi_1,\phi_2) &=& \frac{1}{16aF^2}\int_{-\pi}^\pi\frac{{\rm d}^4\theta}
{(2\pi)^4}\frac{\sum_{k=1}^3\sin^2\theta_k}
{[4+a^2x_{\phi_1}^2/2-\sum_\lambda\cos\theta_\lambda]
[4+a^2x_{\phi_2}^2/2-\sum_\lambda\cos\theta_\lambda]} \nonumber \\
 &+& \frac{1}{16aF^2}\int_{-\pi}^\pi\frac{{\rm d}^3\theta}
{(2\pi)^3}\frac{\sum_{k=1}^3\sin^2\theta_k}
{[3+a^2x_{\phi_1}^2/2-\sum_{k=1}^3\cos\theta_k]
[3+a^2x_{\phi_2}^2/2-\sum_{k=1}^3\cos\theta_k]}, \\
K_2(\phi_1,\phi_2) &=& \frac{1}{8aF^2}\int_{-\pi}^\pi\frac{{\rm d}^4\theta}
{(2\pi)^4}\frac{(1-\cos\theta_4)}
{[4+a^2x_{\phi_1}^2/2-\sum_\lambda\cos\theta_\lambda]
[4+a^2x_{\phi_2}^2/2-\sum_\lambda\cos\theta_\lambda]}, \\
K_3(\phi,\phi) &=& \frac{-{\cal C}^2}{12aF^2}\int_{-\pi}^\pi
\frac{{\rm d}^4\theta}{(2\pi)^4}\frac{(1+a\Delta-\cos\theta_4)\sum_{k=1}^3
\sin^2\theta_k}
{[4+a^2x_{\phi}^2/2-\sum_\lambda\cos\theta_\lambda]^2
[(1+a\Delta)(1-\cos\theta_4)+a^2\Delta^2/2]}, \nonumber \\
\end{eqnarray}
and their coefficients are defined in Table~\ref{tab:sigcoeffs}.

\begin{figure}[tbh]
\vspace*{9mm}
\includegraphics[height=10.0cm]{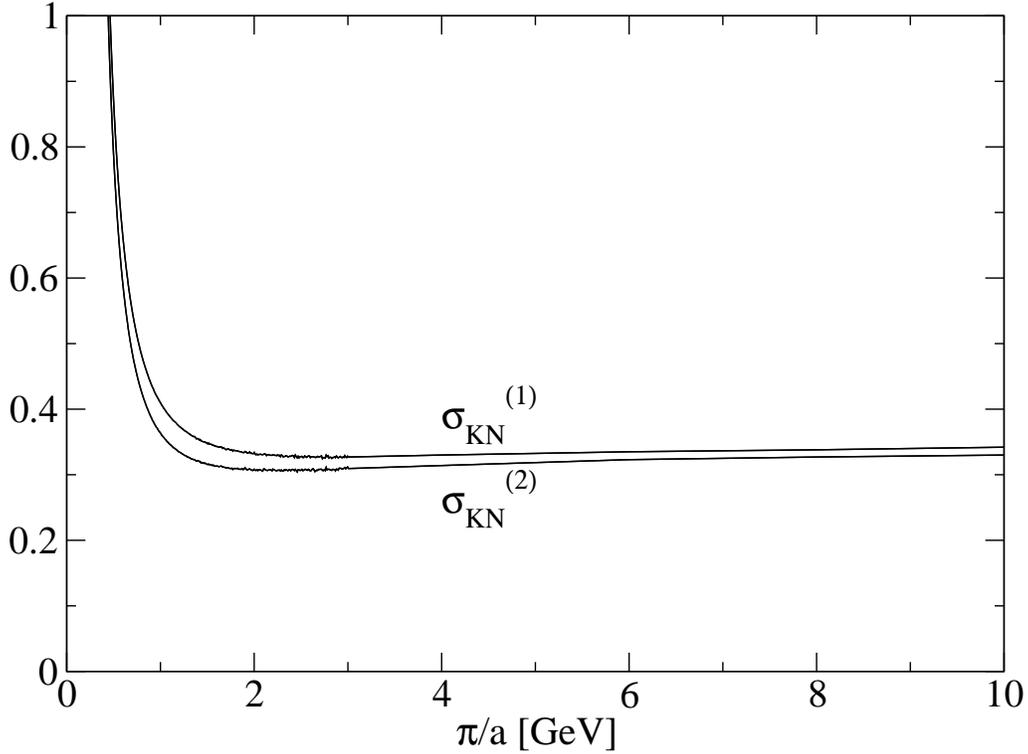}
\caption{Kaon-nucleon sigma terms (in units of GeV) as functions
         of lattice spacing.
         Lagrangian parameters are fixed by requiring the octet baryon
         masses and $\sigma_{\pi N}(0)$ to equal their experimental values
         at all lattice spacings.
        }\label{fig:sigmaKN}
\end{figure}

Following Ref.~\cite{lchpt}, the octet baryon masses and $\sigma_{\pi N}(0)$
are constrained to their physical values at all lattice spacings.
These five observables are functions of five parameters: $m_0$, $b_0$,
$b_{\cal D}$, $b_{\cal F}$ and $\cal D$.
The other axial coupling is ${\cal F} \equiv 1.267-{\cal D}$.
With these parameters fixed, predictions are obtained
for the $KN$ sigma terms as shown in Fig.~\ref{fig:sigmaKN}.
Notice that the $KN$ sigma terms approach their continuum values
very quickly: at $\pi/a=1$~GeV the difference is $O(20\%)$ and
at $\pi/a=6$~GeV the difference is $O(1\%)$.

The momentum dependences of the scalar form factors are
parameter-free, so all lattice spacing effects are exclusively from
loop diagrams.
For definiteness, consider the running of the $\pi N$ sigma term to
the Cheng-Dashen point, $q^2=-2m_\pi^2$ in Euclidean notation.
This is obtained from the
scalar vertex with incoming momentum $q=iQ/a$ where
$Q \equiv (0,0,\sqrt{2}ax_\pi,0)$.
Numerical results can be obtained directly from the integral expression,
\begin{equation}
\sigma_{\pi N}(-2x_\pi^2) - \sigma_{\pi N}(0) = x_\pi^2\sum_{j=1}^3\sum_\phi
       a_j^\phi\left[\tilde{K}_j(\phi)-K_j(\phi,\phi)\right],
\end{equation}
\begin{figure}[tbh]
\vspace*{12mm}
\includegraphics[height=10.0cm]{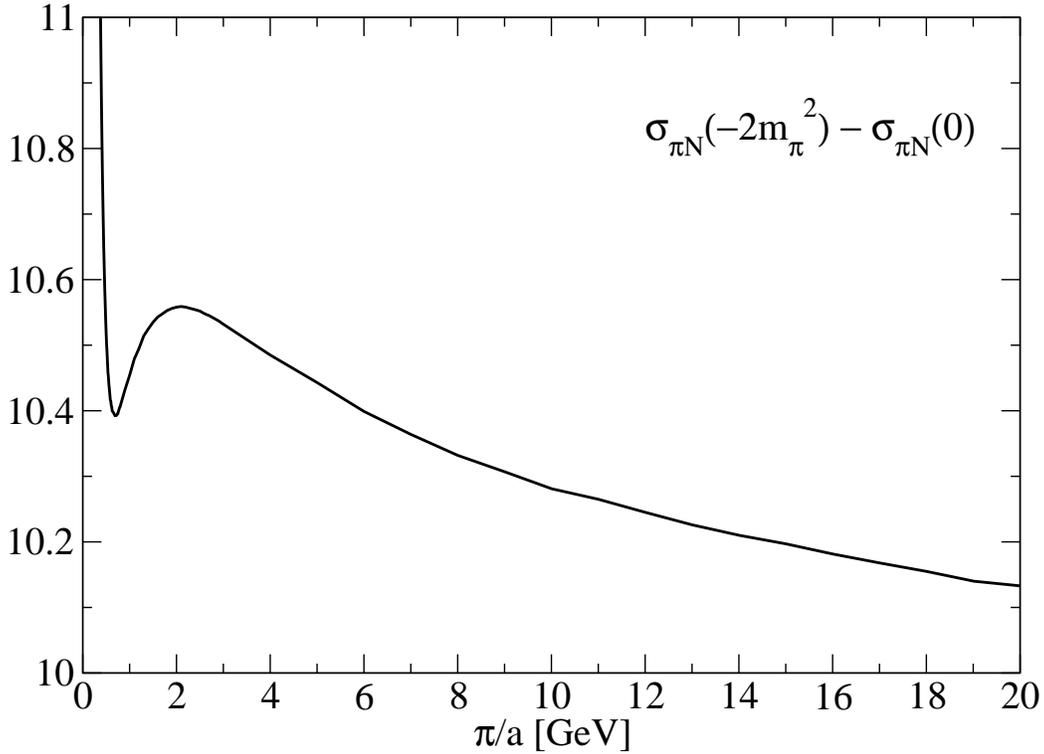}
\caption{The difference (in units of MeV) between the pion-nucleon sigma
         term at the Cheng-Dashen point and at $q^2=0$, as a function
         of lattice spacing.
        }\label{fig:sigmaCD}
\end{figure}
with
\begin{eqnarray}
\tilde K_1(\phi) &=& \frac{1}{16aF^2}\int_{-\pi}^\pi\frac{{\rm d}^4\theta}
{(2\pi)^4}\frac{\sum_{k=1}^3[\sin^2\theta_k\cosh^2Q_k
+\cos^2\theta_k\sinh^2Q_k]}
{[(4+a^2x_{\phi}^2/2-\sum_\lambda\cos\theta_\lambda\cosh Q_k)^2
+(\sum_{k=1}^3\sin\theta_k\sinh Q_k)^2]} \nonumber \\
 &+& \frac{1}{16aF^2}\int_{-\pi}^\pi\frac{{\rm d}^3\theta}
{(2\pi)^3}\frac{\sum_{k=1}^3[\sin^2\theta_k\cosh^2Q_k
+\cos^2\theta_k\sinh^2Q_k]}
{[(3+a^2x_{\phi}^2/2-\sum_{k=1}^3\cos\theta_k\cosh Q_k)^2
+(\sum_{k=1}^3\sin\theta_k\sinh Q_k)^2]}, \nonumber \\ \\
\tilde K_2(\phi) &=& \frac{1}{8aF^2}\int_{-\pi}^\pi\frac{{\rm d}^4\theta}
{(2\pi)^4}\frac{(1-\cos\theta_4)}
{[(4+a^2x_{\phi}^2/2-\sum_\lambda\cos\theta_\lambda\cosh Q_k)^2
+(\sum_{k=1}^3\sin\theta_k\sinh Q_k)^2]}, \\
\tilde K_3(\phi) &=& \frac{-{\cal C}^2}{12aF^2}\int_{-\pi}^\pi
\frac{{\rm d}^4\theta}{(2\pi)^4}\frac{\sum_{k=1}^3
[\sin^2\theta_k\cosh^2Q_k+\cos^2\theta_k\sinh^2Q_k]}
{[(4+a^2x_\phi^2/2-\sum_\lambda\cos\theta_\lambda\cosh Q_\lambda)^2
+(\sum_{k=1}^3\sin\theta_k\sinh Q_k)^2]} \nonumber \\
&& \times \frac{(1+a\Delta-\cos\theta_4)}
{ (1+a\Delta)(1-\cos\theta_4)+a^2\Delta^2/2},
\end{eqnarray}
and the results are plotted in Fig.~\ref{fig:sigmaCD}.
In this case, lattice spacing effects are at the few percent level for
$\pi/a=6$~GeV.

\section*{Acknowledgments}

P.O. is grateful to Wolfram Weise and
the T-39 theory group at Technische Universit\"at M\"unchen
for their support and hospitality during the early stages of this work.
The work was also supported in part by the Deutsche
Forschungsgemeinschaft and the Natural Sciences and Engineering 
Research Council of Canada.

\end{document}